\documentclass[aps,prb,showpacs,twocolumn,reprint,superscriptaddress,longbibliography,showkeys]{revtex4-2}
\usepackage{amsmath}
\usepackage{amssymb}
\usepackage{graphicx}
\usepackage{color}
\usepackage{multirow}
\usepackage{dcolumn}
\usepackage{textcomp}
\usepackage{longtable}
\usepackage{makecell}
\usepackage{booktabs}
\usepackage{array} 
\usepackage{ulem}
\usepackage{appendix}
\usepackage{placeins}

\newcommand{\equalupuparrows}{%
	\begin{array}{c}
		\uparrow \\ \uparrow 
	\end{array}
}

\newcommand{\equaldowndownarrows}{%
	\begin{array}{c}
		\downarrow \\ \downarrow 
	\end{array}
}

\newcommand{\equalupdownarrows}{%
	\begin{array}{c}
		\uparrow \\ \downarrow 
	\end{array}
}

\newcommand{\equaldownuparrows}{%
	\begin{array}{c}
		\downarrow \\ \uparrow 
	\end{array}
}

\newcommand{\equaldownupuparrows}{%
	\begin{array}{c}
		\uparrow \\ \uparrow \\ \downarrow 
	\end{array}
}

\newcommand{\equaldowndownuparrows}{%
	\begin{array}{c}
		\uparrow \\ \downarrow \\ \downarrow
\end{array}
}

\newcommand{\equalupupuparrows}{%
	\begin{array}{c}
		\uparrow \\ \uparrow \\ \uparrow 
	\end{array}
}

\newcommand{\equaldowndowndownarrows}{%
	\begin{array}{c}
		\downarrow \\ \downarrow \\ \downarrow 
	\end{array}
}

\newcommand{\equalupdownuparrows}{%
	\begin{array}{c}
		\uparrow \\ \downarrow \\ \uparrow
	\end{array}
}

\newcommand{\equaldownupdownarrows}{%
	\begin{array}{c}
		\downarrow \\ \uparrow \\ \downarrow 
	\end{array}
}

\newcommand{\equalupupdownarrows}{%
	\begin{array}{c}
		\downarrow \\ \uparrow \\ \uparrow
	\end{array}
}

\newcommand{\equalupdowndownarrows}{%
	\begin{array}{c}
		\downarrow \\ \downarrow \\ \uparrow 
	\end{array}
}

\newcommand{\equalupupupuparrows}{%
	\begin{array}{c}
		\uparrow \\ \uparrow \\ \uparrow \\ \uparrow
	\end{array}
}

\newcommand{\equaldowndowndowndownarrows}{%
	\begin{array}{c}
		\downarrow \\ \downarrow \\ \downarrow \\ \downarrow
	\end{array}
}

\newcommand{\equaldowndownupuparrows}{%
	\begin{array}{c}
		\uparrow \\ \uparrow \\ \downarrow \\ \downarrow
	\end{array}
}

\newcommand{\equaldownupupuparrows}{%
	\begin{array}{c}
		\uparrow \\ \uparrow \\ \uparrow \\ \downarrow
	\end{array}
}

\newcommand{\equalupupupdownarrows}{%
	\begin{array}{c}
		\downarrow \\ \uparrow \\ \uparrow \\ \uparrow
	\end{array}
}

\newcommand{\equalupupdowndownarrows}{%
	\begin{array}{c}
		\downarrow \\ \downarrow \\ \uparrow \\ \uparrow
	\end{array}
}

\newcommand{\equalupdowndownuparrows}{%
	\begin{array}{c}
		\uparrow \\ \downarrow \\ \downarrow \\ \uparrow
	\end{array}
}

\newcommand{\equaldownupdowndownarrows}{%
	\begin{array}{c}
		\downarrow \\ \downarrow \\ \uparrow \\ \downarrow
	\end{array}
}

\newcommand{\equaldowndownupdownarrows}{%
	\begin{array}{c}
		\downarrow \\ \uparrow \\ \downarrow \\ \downarrow
	\end{array}
}

\newcommand{\equaldownupupdownarrows}{%
	\begin{array}{c}
		\downarrow \\ \uparrow \\ \uparrow \\ \downarrow
	\end{array}
}

\newcommand{\equaldownupdownuparrows}{%
	\begin{array}{c}
		\downarrow \\ \uparrow \\ \downarrow \\ \uparrow
	\end{array}
}

\newcommand{\equalupdownupdownarrows}{%
	\begin{array}{c}
		\uparrow \\ \downarrow \\ \uparrow \\ \downarrow
	\end{array}
}

\newcommand{\equalupdownupuparrows}{%
	\begin{array}{c}
		\uparrow \\ \uparrow \\ \downarrow \\ \uparrow
	\end{array}
}

\newcommand{\equalupupdownuparrows}{%
	\begin{array}{c}
		\uparrow \\ \downarrow \\ \uparrow \\ \uparrow
	\end{array}
}

\begin{document}

%\title{Interface-induced antiferroelectric-ferroelectric transition in layered two dimensional materials}

\title{Ferroelectrically tunable topological phase transition in In$_2$Se$_3$ thin films}

\author {Zhiqiang Tian}
\affiliation{Key Laboratory for Matter Microstructure and Function of Hunan Province,
Key Laboratory of Low-Dimensional Quantum Structures and Quantum Control of Ministry of Education, School of Physics and Electronics, Hunan Normal University, Changsha 410081, China}

\author {Ziming Zhu}
\affiliation{Key Laboratory for Matter Microstructure and Function of Hunan Province,
Key Laboratory of Low-Dimensional Quantum Structures and Quantum Control of Ministry of Education, School of Physics and Electronics, Hunan Normal University, Changsha 410081, China}

\author {Jiang Zeng}
\affiliation{School of Physics and Electronics, Hunan University, Changsha 410082, People’s Republic of China}

\author {Chao-Fei Liu}
\affiliation{School of Science, Jiangxi University of Science and Technology, Ganzhou 341000, China}

\author {Yurong Yang}
\affiliation{National Laboratory of Solid State Microstructures and Collaborative Innovation Center of Advanced Microstructures, Nanjing University, Nanjing 210093, China}

\author {Anlian Pan}
\affiliation{Key Laboratory for Micro-Nano Physics and Technology of Hunan Province, College of Materials Science and Engineering, Hunan University, Changsha 410082, China}

\author {Mingxing Chen}
\email{mxchen@hunnu.edu.cn}
\affiliation{Key Laboratory for Matter Microstructure and Function of Hunan Province,
Key Laboratory of Low-Dimensional Quantum Structures and Quantum Control of Ministry of Education, School of Physics and Electronics, Hunan Normal University, Changsha 410081, China}
\affiliation{State Key Laboratory of Powder Metallurgy, Central South University,  Changsha 410083, China}

\date{\today}

\begin{abstract}
Materials with ferroelectrically switchable topological properties are of interest for both fundamental physics and practical applications. Using first-principles calculations, we find that stacking ferroelectric $\alpha$-In$_2$Se$_3$ monolayers into a bilayer leads to polarization-dependent band structures, which yields polarization-dependent topological properties. Specifically, we find that the states with interlayer ferroelectric couplings are quantum spin Hall insulators, while those with antiferroelectric polarizations are normal insulators.  We further find that In$_2$Se$_3$ trilayer and quadlayer exhibit nontrivial band topology as long as in the structure the ferroelectric In$_2$Se$_3$ bilayer is antiferroelectrically coupled to In$_2$Se$_3$ monolayers or other ferroelectric In$_2$Se$_3$ bilayer. Otherwise the system is topologically trivial. The reason is that near the Fermi level the band structure of the ferroelectric In$_2$Se$_3$ bilayer has to be maintained for the nontrivial band topology. This feature can be used to design nontrivial band topology for the thicker films by a proper combination of the interlayer polarization couplings. The topological properties can be ferroelectrically tunable using the dipole locking effect.  Our study reveals switchable band topology in a family of natural ferroelectrics,  which provide a platform for designing new functional devices.  
\end{abstract}


\maketitle
%\section{INTRODUCTION}
Ferroelectrics have attracted great attention for applications in manufacturing low power and high speed memory devices \cite{Scott2007}. While topological insulators emerging in the past decade are promising in spintronic devices and realizing Majorana Fermion for quantum computing due to the gapless surface/edge states with spin-momentum locking \cite{RMP2010, RMP2011}.  Physically, band topology and ferroelectricity are seemly uncorrelated since they have different origins. Materials with nontrivial band topology such as topological insulators have the electronic origin, for which spin-orbit coupling plays the role \cite{RMP2010, RMP2011}. Whereas most of ferroelectrics such as ferroelectric (FE) perovskites have the ionic origin which is caused by displacement of certain atoms \cite{Cohen1992}.  Moreover, these two families of materials are also distinct in band gap that topological insulators usually have a band gap less than 1.0 eV \cite{Yan2012, Vergniory2019,Zhang2019,Tang2019}. However, ferroelectrics have much larger band gaps than the topological insulators.  Combining these two into one material not only is of interest for fundamental physics, but also provides opportunities for designing new functional devices. 

There have been a few attempts to explore materials that simultaneously exhibit ferroelectricity and nontrivial band topology.  Density functional theory (DFT) calculations reveal that SnTe experiences a topological phase transition, i.e.,  changing from a topological crystalline insulator into a topological insulator as it undergoes a phase transition from the room-temperature cubic phase into low-temperature ferroelectric structure \cite{Plekhanov2014}.  Recently, DFT calculations show that strain can drive a cubic perovskite CsPbI$_3$ into a noncentrosymmetric structure with ferroelectricity and can be turned into a topological insulator from a normal insulator under appropriate strains \cite{Liu2016}.  It is also revealed that several antiferroelectric (AFE) alkali-MgBi orthorhombic members are topological insulators. In these materials, the polar and antipolar states have distinct topological properties, i.e., topologically trivial and nontrivial phases \cite{Monserrat2017}. This feature allows for an electric field control of topological phase transition as the system switches between the antipolar and polar states. 

The rise of two-dimensional (2D) FE materials \cite{Chang2016, Fei2016, Higashitarumizu2020,Ding2017,Zhou2017, Cui2018, ZengH2018, ZengH2019, CIPS2016} further stimulates efforts toward ferroelectric control of topological phase transitions in layered 2D materials.  In particular, CuInP$_2$S$_6$ thin films and $\alpha$-In$_2$Se$_3$ monolayer possess out-of-plane polarizations \cite{Ding2017, CIPS2016}, which are good for ferroelectric tuning electronic properties of surface layers in interface structures\cite{Gong2019,Zhuz2019,Sun2020,Chen2020,HuDuan2020, Xue2020,FengD2021, Yu2023}.  In the heterostructures of Bi(111)-bilayer/$\alpha$-In$_2$Se$_3$ and Sb(111)-bilayer/$\alpha$-In$_2$Se$_3$,  the surface overlayers experience a transition from a topological insulator to a normal insulator as the reversal of the electric polarization of $\alpha$-In$_2$Se$_3$ \cite{Bai2020,Zhang2021}.  It is predicted that nontrivial band topology and ferroelectricity coexist in a bismuth monolayer functionalized by an organic molecule \cite{Kou2018}.  Reversing the polarization can result in a reversal of the spin texture of electronic bands, but remains the band topology unchanged. Recently, Huang \textit{et al} propose to design the 2D topological insulators with ferroelectrically tunable topological properties using the heterostructure of In$_2$S$_3$/In$_2$Se$_3$ as the illustration. This system called type-II 2D ferroelectric topological insulator is a normal insulator when it has the up polarizations. Then it becomes a topological insulator as all the polarizations are reversed \cite{2DFETI}. Despite these achievements, natural 2D single-phase ferroelectrics exhibiting tunable topological properties are still rare.

In this work, we investigate the electronic and topological properties of $\alpha$-In$_2$Se$_3$ thin films using first-principles calculations. We find that its monolayer is a normal insulator, of which the stacking gives rise to polarization dependent topological properties for the bilayer. Specifically, the states with interlayer FE orderings exhibit nontrivial band topology, whereas the others are topologically trivial. Such behavior is caused by the polarization dependent interlayer band hybridization. The same physics can be applied to understand the topological properties of the trilayer and quadlayer, in which the FE bilayer serves as a building block. As a result, the polarization states of an In$_2$Se$_3$ thin film can be categorized into two groups: a quantum spin Hall insulator and a normal insulator, which depends whether near the Fermi level the characteristics of the FE bilayer is preserved. The topological properties can be ferroelectrically tunable by making use of the dipole locking effect. Therefore,  a ferroelectrically switchable topological transition can be achieved in $\alpha$-In$_2$Se$_3$ thin films, which has potential applications in new functional electronic devices. 

%\section{COMPUTATIONAL DETAILS}
Our DFT calculations were performed using the Vienna Ab initio Simulation Package (VASP) \cite{kresse1996}. The pseudopotentials were constructed by the projector augmented wave method \cite{blochl1994,kresse1999}. The exchange-correlation functional is parametrized using the Perdew–Burke–Ernzerhof (PBE) formalism in the generalized gradient approximation~\cite{PBE1996}. An 12 $\times$ 12 $\times$ 1 and 24 $\times$ 24 $\times$ 1 $ \Gamma $-centered k-mesh were used to sample the 2D Brillouin zone for structural relaxation and electronic structure calculations, respectively. An energy cutoff of 400 eV was used for the plane-waves for all calculations. A 20 \AA vacuum region was used between adjacent plates to avoid the artificial interaction between neighboring periodic images. Since the systems are layered systems, van der Waals (vdW) dispersion forces between the layers were accounted for through the DFT-D3 \cite{DFT-D3}. The structures were fully relaxed until the residual force on each atom was less than 0.01 eV/\AA. A saw-like self-consistent dipole layer was placed in the middle of the vacuum region to adjust the misalignment between the vacuum levels on the different sides of the film due to the intrinsic electric polarization. The kinetic pathways of transitions between different polarization states are calculated using the climbing image nudge elastic band (CI-NEB) method \cite{Henkelman2000,henkelman2000improved}. The topological properties calculations were carried out using the WANNIER90 \cite{wannier90} and WannierTools package \cite{WU2017}.

\begin{figure}[t]
  \includegraphics[width=.95\linewidth]{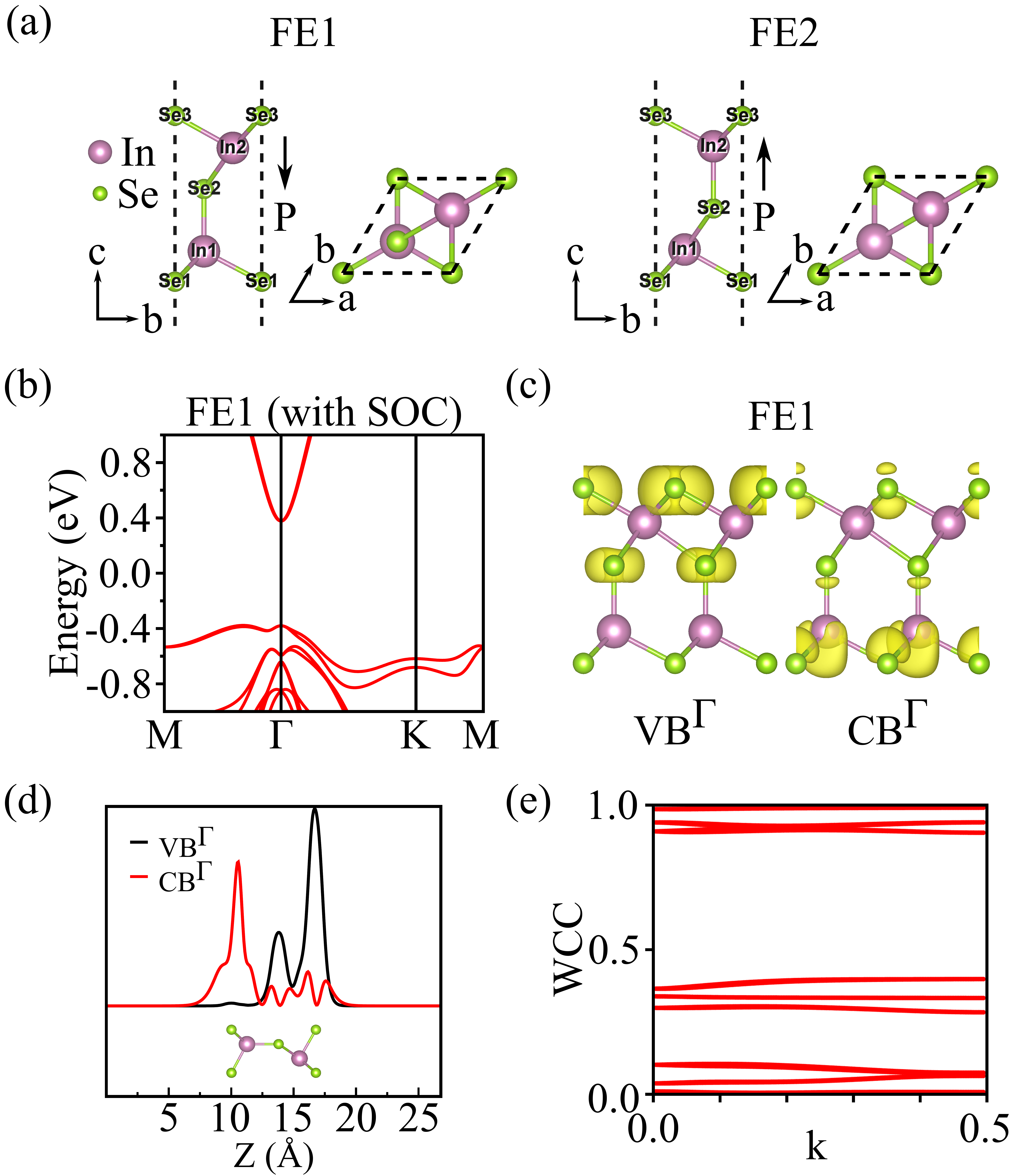}
  \caption{Electronic structure of In$_2$Se$_3$-1L. (a) Geometric structure of In$_2$Se$_3$-1L with different polarization states. The atoms are labelled by numbers. The polarizations are shown by arrows.  The state with downward polarizations is denoted as FE1, whereas the one with upward polarizations is referred to as FE2. (b) Band structure with spin-orbit coupling. (c) Charge density distribution of the valence band and conduction band at $\Gamma$, i.e., VB$^\Gamma$ and CB$^\Gamma$. (d) Wave functions at VB$^\Gamma$ and CB$^\Gamma$. (e) Evolution of WCC for In$_2$Se$_3$-1L.}
 \label{fig1}
\end{figure}

%\section{RESULTS AND DISCUSSIONS}
%\subsection{Characteristics of bands for $\alpha$-In$_2$Se$_3$ monolayer}
%
We begin by discussing the band structure and corresponding band topology for the ferroelectric $\alpha$-In$_2$Se$_3$ monolayer (In$_2$Se$_3$-1L), whose geometric structure is shown in Fig.~\ref{fig1}(a). All the Se atoms in the unit cell are inequivalent due to the charge polarization. Therefore, we label the Se atoms as Se1, Se2, and Se3, respectively. Likewise, the two In atoms are denoted as In1 and In2, respectively. The simultaneous charge polarization is due to the motion of Se2.  We label the two ferroelectric states with opposite polarizations as FE1 and FE2, respectively. Fig.~\ref{fig1}(b) shows its band structure with the inclusion of spin-orbit coupling (SOC). The one without SOC is pretty much similar and agrees well with previous studies \cite{Ding2017}.  A comparison of the band structures from the two types of calculations finds that the band gap is reduced by only about 8 meV when the SOC is included. The similarity in the band structure between them implies that In$_2$Se$_3$-1L may have trivial band topology. We further analyze the characteristics of the bands for the FE1 state (the one with downward polarization).  Fig.~\ref{fig1}(c) shows the charge density distribution of the valence and conduction bands at $\Gamma$, which will be hereafter referred to as VB$^\Gamma$ and CB$^\Gamma$.  One can see that VB$^\Gamma$  is mainly contributed by Se2 and Se3 atoms, whereas CB$^\Gamma$ is dominated by orbitals of Se$_1$. Fig.~\ref{fig1}(d) shows that the wavefunction of VB$^\Gamma$ has a much shorter decay length than that of CB$^\Gamma$. Therefore, the wavefunction of VB$^\Gamma$ is more localized than that of CB$^\Gamma$. As Se2 moves downward so that the system transforms into the FE2 state, the charge distributions are also reversed correspondingly.  Fig.~\ref{fig1}(e) depicts the evolution of the Wannier charge center (WCC) for In$_2$Se$_3$-1L, which indicates that it is topologically trivial.  We have also performed calculations of edge states for the system (not shown), which again confirms that topologically In$_2$Se$_3$-1L is a normal insulator.

\begin{figure}[t]
 \includegraphics[width=.95\linewidth]{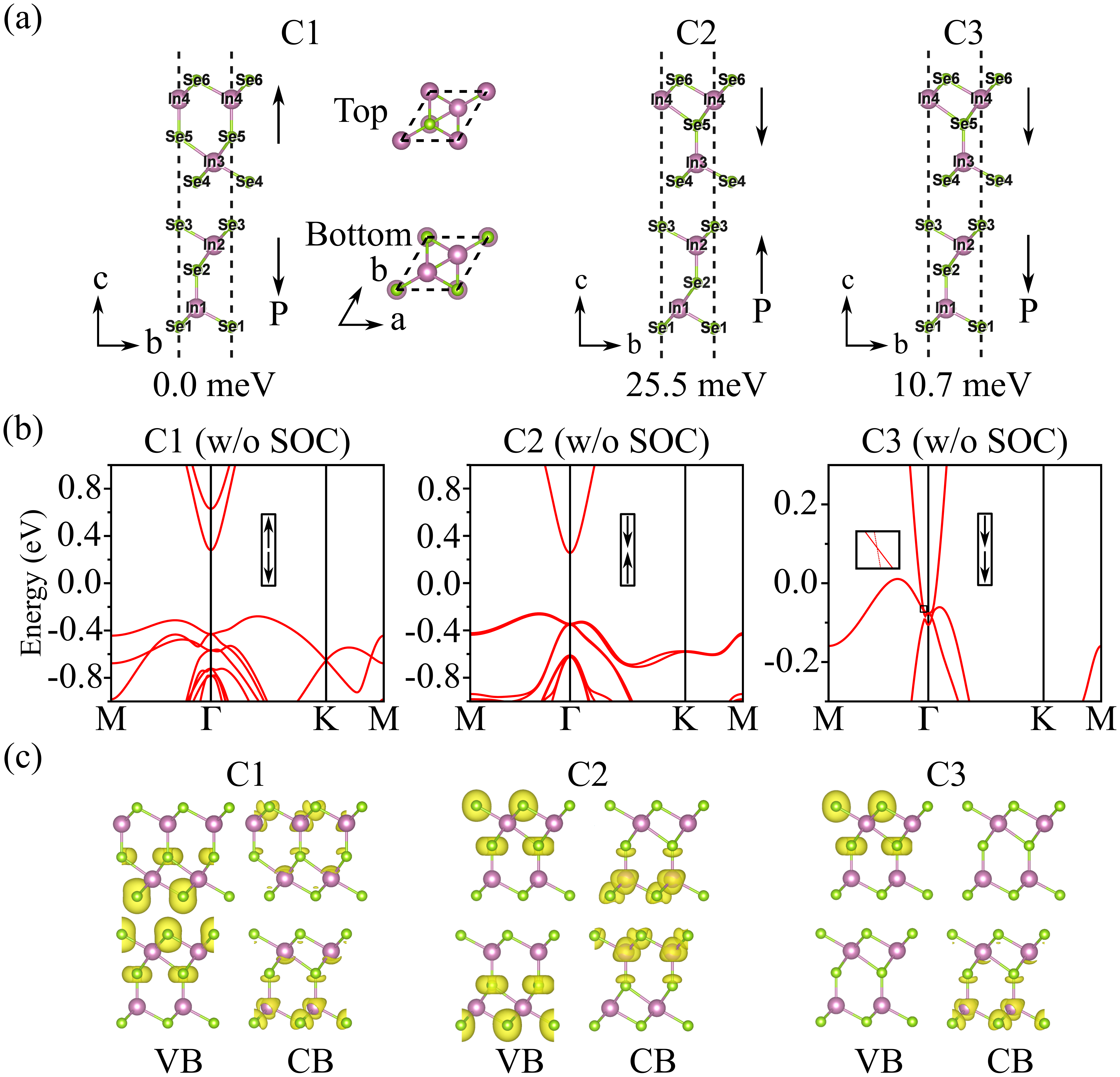}
  \caption{Polarization dependence of the band structure for In$_2$Se$_3$-2L. (a) Polarization states of In$_2$Se$_3$-2L. The total energy for each configuration is given below the structure. (b) Band structures for the states shown in (a).  (c) Wavefunctions of the valence and conduction bands.} 
 \label{fig2}
\end{figure}

%\subsection{$\alpha$-In$_2$Se$_3$ bilayer}
%
There are two types of stacking orders when placing two In$_2$Se$_3$-1Ls together into a bilayer (In$_2$Se$_3$-2L), i.e., the R- and H-stackings. In our calculations, they are in the R-stacking, which is also the stacking order in the bulk phase.  There are four configurations for a bilayer by considering different interlayer polarization orderings. Three possible stacking orders for all polarization configurations have been considered (see Appendix A for more details). The most stable stacking for each polarization configuration is shown in Fig.~\ref{fig2}a, which is consistent with previous  studies~\cite{Cui2018,Liujunming2020,Yan2022}. Among them, two have interlayer AFE oderings, which are denoted as C1, and C2. The polarizations in them are in the ways of tail-to-tail and head-to-head, respectively. The other two have the interlayer FE ordering with opposite total polarizations, which are named C3 and C4 (not shown), respectively.   Like Ref.~\onlinecite{Monserrat2017},  C1 and C2 may be referred to as antipolar phases, while C3 and C4 are polar phases. Among the four configurations, C1 has the lowest energy among all the configurations.  while C3 and C4 can be related by the inversion symmetry. Therefore, they are energetically degenerate.  

Figure~\ref{fig2}(b) shows the band structures without SOC for the polarization states shown above.  There are significant differences between them. First, the states with interlayer AFE couplings, i.e., C1 and C2, are semiconductors. Whereas, those with the FE coupling are semimetals. Moreover, the band gaps between the AFE states are also different, which are about 0.6 and 0.4 eV for C1 and C2, respectively.  These values are in good agreement with those reported Ref.\onlinecite{DingPRL2021} . The polarization dependence of the band structures is caused by the polarization-dependent band hybridizations, which are related to the characteristics of the wavefunctions as shown in Fig.~\ref{fig1}(c). 
For C1,  the valence band is mainly contributed by the interfacial Se atoms.  Whereas the conduction band is dominated by contributions of the outmost Se atoms, for which the wavefunctions are delocalized. For C2, the situation is on the opposite [see Fig.~\ref{fig2}(c)].  
So, one can expect that interlayer hybridizations between the valence bands of the two layers for C1 are stronger than those for C2.  
Note that the wavefunction of VB$^\Gamma$ is localized and the distance between the outmost atoms of the two layers is about 17 \AA. Therefore, the layer interaction has a minor effect on the valence band for C2. This effect can also be seen in that the valence band for C2 is closely similar to that of the monolayer.  The trend for the conduction bands is the opposite, that is, the interlayer hybridizations are strong for C2 and weak for C1.
Therefore, the band splitting in the conduction band due to stacking for C2 is larger than that for C1.  

\begin{figure*}[t]
	\includegraphics[width=.9\linewidth]{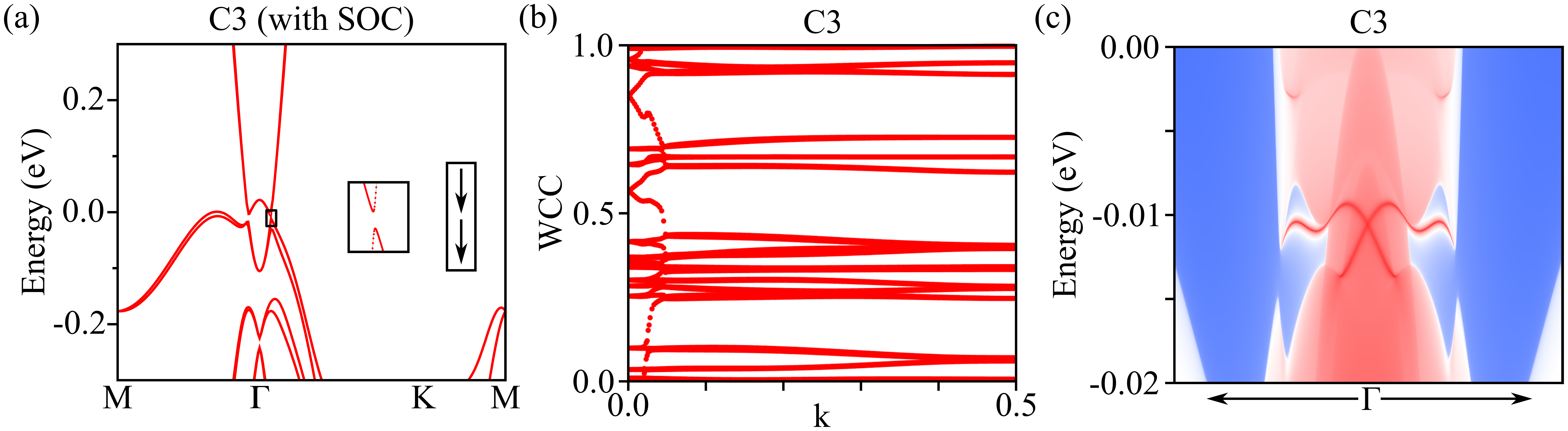}
	\caption{Band topology of FE In$_2$Se$_3$-2L, i.e., configuration C3. (a) Band structure with SOC. (b) Evolution of WCC. (c) Boundary states of a zigzag edge. } 
	\label{fig3}
\end{figure*}

The FE state is a semimetal, which can be seen from the band structure for C3. This result consistent with previous studies can be understood that the built-in electric field shifts the bands of the two monolayers.  The relative shifting of the bands for the two layers leads to negligible hybridizations between the valence bands as well as those between the conduction bands [see Fig.~\ref{fig2}(c)]. Instead, the valence band of the top In$_2$Se$_3$ layer hybridizes with the conduction band of the bottom layer, which results in charge transfer between the layers [see Appendix Fig.~\ref{fig7}(a)]  We recall that the VB$^\Gamma$ of the top layer is mainly contributed by Se3 and the CB$^\Gamma$ of the bottom layer is dominated by contributions of Se6. So,  these atoms are geometrically far from each other, which means that the hoppings between their orbitals are small. As a result, the induced band gap due to the interlayer band 
hybridizations is small, which is about 10 meV along $\Gamma$-K [Fig.~\ref{fig2}(c)].   
Note that there is a band crossing near the Fermi level at a $k$-point along $\Gamma$-M due to symmetry protection, pointing to novel electronic properties for this system, which will be discussed below. 

%\begin{figure*}[t]
%  \includegraphics[width=.9\linewidth]{Fig3.pdf}
%  \caption{Band topology of FE In$_2$Se$_3$-2L, i.e., configuration C3. (a) Band structure with SOC. (b) Evolution of WCC for configurations. (c) Boundary states of a zigzag edge for C3. } 
% \label{fig3}
%\end{figure*}

Figure~\ref{fig3}(a) shows the band structure with SOC for C3.  One prominent feature is that a SOC gap of about 10 meV appears at the $k$-point along $\Gamma$-M where a band crossing happens for the calculations without SOC. We have calculated the WCC for both the AFE (C1) and FE states (C3), for which the results are shown in Fig.~\ref{fig3}(b) (results for C1 are not shown). We obtain Z2 = 0 for C1 and C2 and Z2 = 1 for C3 and C4, respectively. We have further performed calculations of a semi-infinite ribbon for C3, from which one can see gapless edges [Fig.~\ref{fig3}(c)]. Therefore, C3 and C4 have a nontrivial band topology, whereas C1 and C2 are topologically trivial.  The polarization-dependent band structure and band topology for In$_2$Se$_3$-2L can be used to understand the results for the multilayers. 

%\subsection{$\alpha$-In$_2$Se$_3$ multilayers}

\begin{figure}[t]
 \includegraphics[width=.95\linewidth]{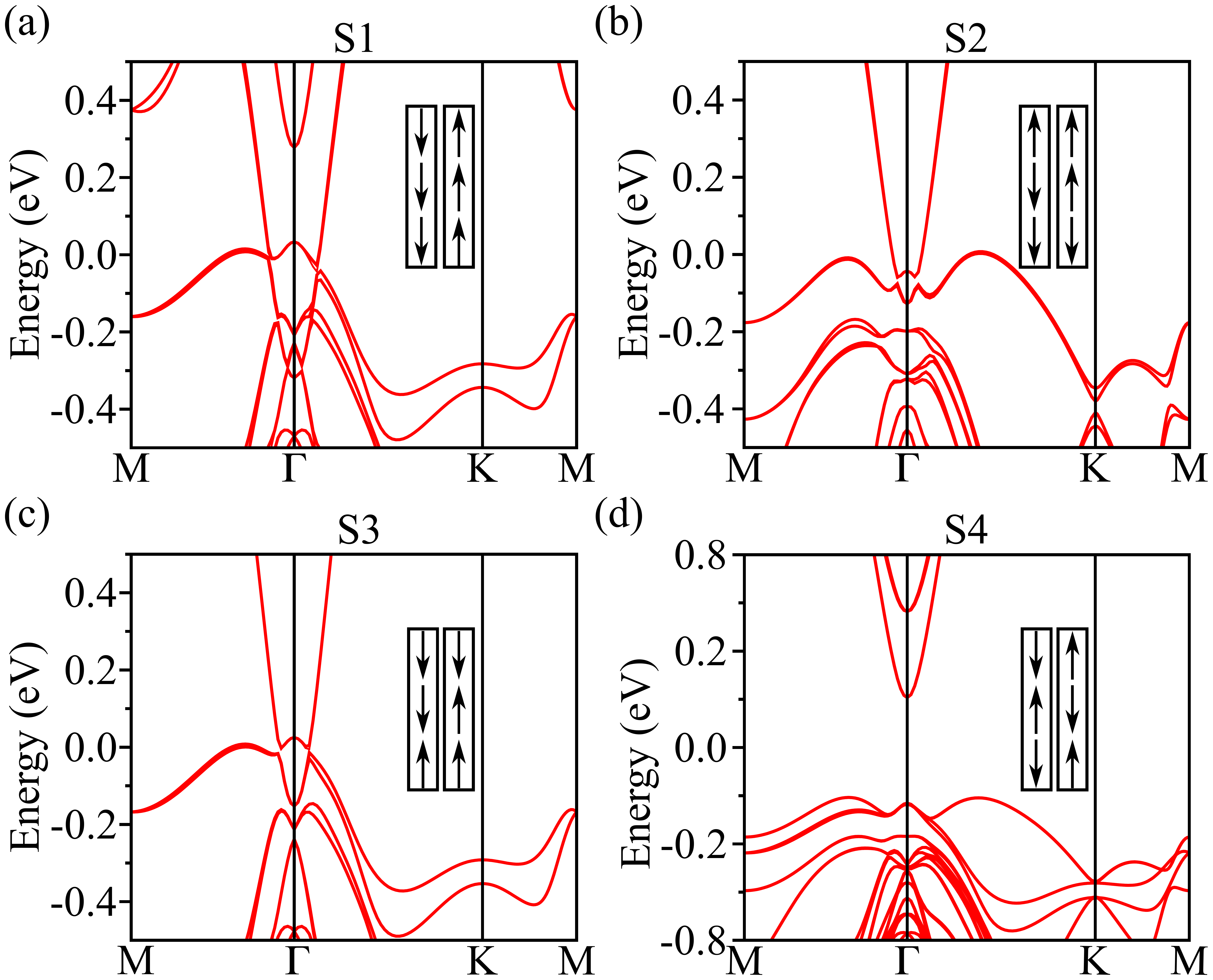}
  \caption{Band structures with SOC for In$_2$Se$_3$-3L in four states.  The insets show equivalent polarization configurations. } 
 \label{fig4}
\end{figure}

Figure~\ref{fig4} shows the band structures for an In$_2$Se$_3$ trilayer (In$_2$Se$_3$-3L) in different polarization states. The equivalent polarization configurations for each state are shown in the insets. For the FE state [S1 in Fig.~\ref{fig4}(a)] the conduction band of the bottom In$_2$Se$_3$ layer and the valence band of the top layer has a larger overlap than that for the bilayer.  Both the valence and the conduction bands are crossing the Fermi level, exhibiting a metallic band structure. 
A close inspection of the bands near the crossing point finds that like the FE In$_2$Se$_3$-2L there is still a gap opening at a $k$-point along $\Gamma$-K. However, it is much smaller than that for In$_2$Se$_3$-2L.  
The reason is that the hybridizations between these states are extremely small. We performed layer projection of the band structure for this system, which is somehow similar to that for the FE In$_2$Se$_3$-2L near the Fermi level [see Appendix Fig.~\ref{fig7}(b)].  However, one can also see visible changes in the bands near -0.2 eV at $\Gamma$ due to the presence of the third layer. The valence band is mainly contributed by the topmost Se atoms in the top layer and the lowermost
 Se atoms in the bottom layer. Therefore, these atoms are much further from each other than in the FE In$_2$Se$_3$-2L, which negligible band hybridizations.  Moreover, we note that 
For the ferrielectric states S2 and S3,  the band profile near the Fermi level has a great similarity with that for the FE In$_2$Se$_3$-2L.  A comparison of the band structure with those for In$_2$Se$_3$-2L finds that they look like a superimpose of the band structures of C2 and C3 for In$_2$Se$_3$-2L.  This feature can be understood since geometrically S2 (S3) for In$_2$Se$_3$-3L is a combination of C1 (C2) and C3 for In$_2$Se$_3$-2L.  
S4 has a different band structure from those of S2 and S3 in that its band gap is much larger than that for the laters. The reason is that there are only AFE interlayer couplings in this state. Namely, only C1 and C2 configurations for In$_2$Se$_3$-2L are involved in forming S4.  The combination of C1 and C2 gives a band structure with large splittings in both the valence and conduction bands for S4.  

\begin{figure*}[t]
 \includegraphics[width=.95\linewidth]{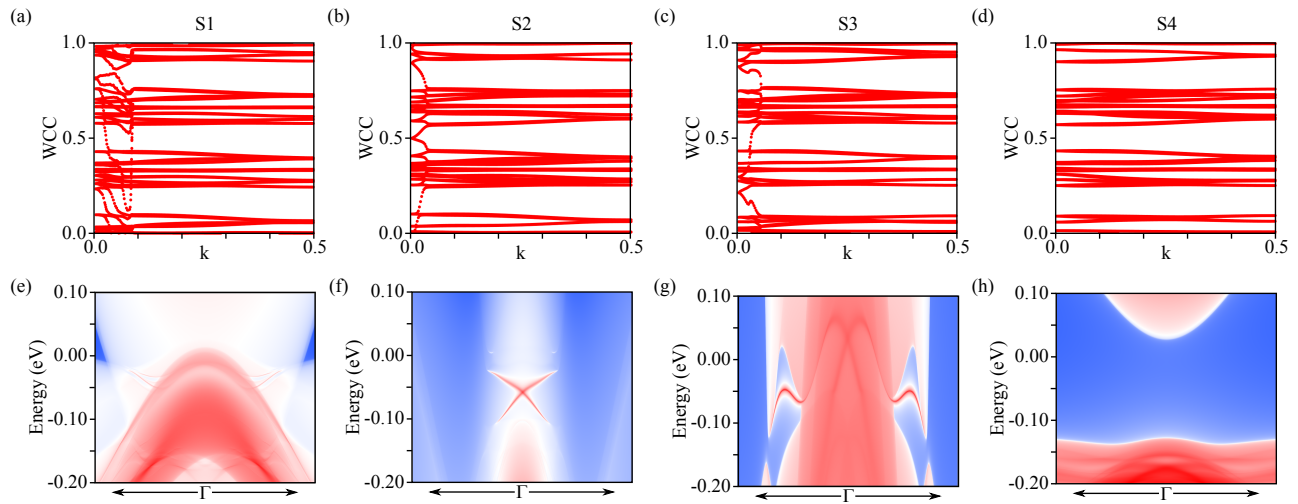}
  \caption{(a-d) Evolution of WCC for In$_2$Se$_3$-3L in the polarization states of S1, S2, S3, and S4, respectively. (f-h) Corresponding boundary states of a zigzag edge for each of them. } 
 \label{fig5}
\end{figure*}

We now discuss the band topology of In$_2$Se$_3$-3L.  Fig.~\ref{fig5} shows the evolution of the WCC for the polarization states.  We obtain Z2 = 0 for S1 and S4 and Z2 = 1 for S2 and S3, which suggest that S2 and S3 are topologically nontrivial. The reason for that S1 has Z2 = 0 may be due to that the presence of the third In$_2$Se$_3$ changes the band characteristics of the FE In$_2$Se$_3$-2L when it is ferroelectrically coupled to the FE In$_2$Se$_3$-2L. We have further performed calculations of the boundary states for both zigzag and armchair edges. The results for the zigzag edges are depicted in Figs.~\ref{fig5}(e-h). One can see that there are gapless edge states for S2 and S3, which confirms the nontrivial band topology.  The calculations of the armchair edges show the same trend. We have also investigated the electronic structure and band topology for an In$_2$Se$_3$ quadlayer (In$_2$Se$_3$-4L), for which the results of the band structure and the WCC are shown in Appendix Fig.~\ref{fig8}.  Table~\ref{table1} summarizes the value of Z2 for all the bilayer, trilayer, and quadlayers.  Our results reveal that $\alpha$-In$_2$Se$_3$ thin films have nontrivial band topology as long as near the Fermi level characteristics of the electronic bands of the FE In$_2$Se$_3$-2L are maintained. This type of band structure can be achieved by properly stacking the FE In$_2$Se$_3$-2L with other In$_2$Se$_3$ layers. Specifically, an In$_2$Se$_3$ thin film is expected to be a quantum spin Hall insulator in the case that the FE In$_2$Se$_3$-2L has antiferroelectric interlayer couplings with other  In$_2$Se$_3$ layers.  Therefore, the FE In$_2$Se$_3$-2L can be used as a building block for designing topological materials.  Moreover, we further investigated the effect of a vdW substrate on the topological states.  We used a h-BN monolayer as the substrate in our calculations, for which the results are shown in Appendix Fig.~\ref{fig9}. One can see that the nontrivial band topology for In$_2$Se$_3$-2L is robust against the perturbation of a vdW substrate. We expect that this trend can be observed for thicker films, e.g., In$_2$Se$_3$-3L and In$_2$Se$_3$-4L.  The nontrivial band topology of In$_2$Se$_3$ thin films is further confirmed by our Hyed-Scuseria-Ernzerhof (HSE) hybrid exchange-correlation functional\cite{HSE062006} calculations (Appendix E).

\begin{table}
	\setlength{\tabcolsep}{15pt}
	\renewcommand{\arraystretch}{1.2}
	%\centering
	\caption {Values of Z2 for the bilayer, trilayer, and quadlayer of $\alpha$-In$_2$Se$_3$ in different polarization states. }
	%\begin{ruledtabular}
	\label{table1} 
	\begin{tabular}{c|c|c}
		\toprule[0.7 pt] 
		\bottomrule[0.7 pt]
    	Systems & Polarization states & Z2 \\
    	%\hline
    	%\textrm{Systems}&
    	%\textrm{Polarization states}&
    	%\textrm{Z2}\\
    	\hline
    	\multirow{3}{*}{\centering In$_2$Se$_3$-2L} & $\equalupuparrows$ $\equaldowndownarrows$  & 1\\
	    \cline{2-3}
    	&  $\equalupdownarrows$ $\equaldownuparrows$ & 0  \\
    	\hline
    	%\colrule
    	\multirow{4}{*}{\centering In$_2$Se$_3$-3L} & $\equalupupdownarrows$ $\equalupdowndownarrows$ $\equaldowndownuparrows$ $\equaldownupuparrows$  & 1\\
    	\cline{2-3}
    	&  $\equalupupuparrows$ $\equalupdownuparrows$ $\equaldownupdownarrows$ $\equaldowndowndownarrows$ & 0  \\
    	\hline
    	\multirow{5}{*}{\centering In$_2$Se$_3$-4L}  &  $\equalupupdownuparrows$ $\equalupupdowndownarrows$ $\equalupdownupuparrows$ $\equalupdowndownuparrows$  $\equaldowndownupuparrows$ $\equaldownupupdownarrows$ $\equaldowndownupdownarrows$ $\equaldownupdowndownarrows$ & 1 \\
    	\cline{2-3}
    	&  $\equalupupupuparrows$ $\equalupupupdownarrows$ $\equalupupupuparrows$ $\equaldownupdownuparrows$  $\equalupdownupdownarrows$ $\equalupdownupdownarrows$ $\equaldownupupuparrows$ $\equaldowndowndowndownarrows$ & 0\\
		
		\toprule[0.7 pt] 
		\toprule[0.7 pt]
	\end{tabular}
%\end{ruledtabular}
\end{table}

The three states with distinct topological properties for the In$_2$Se$_3$-2L are ferroelectrically tunable. This can be achieved by suitable perpendicular electric fields. Below we show that there is a layer selective flipping mechanism for this system due to the interlayer AFE coupling for the ground state. This mechanism can be derived from the energy barriers for the transforming between them. The energy barrier for the FE state (C3) transforming into the AFE state with tail-to-tail configuration (C1) is about 33 meV. This energy barrier, named as E$_B^1$, is lower than the one for the path from C3 to C2. It is also lower than the one for completely flipping the polarizations of C3, i.e., from C3 to C4.  The energy barrier for the pathway from C1 to C4 is about 43 meV (E$_B^2$).  As a result,  the energy barriers for flipping the top and bottom layers are different. The higher the barrier height, the larger the electric field is required for the transforming. Supposing that the system is in C3 (this can be achieved by poling), an opposite electric field that only overcoming E$_B^1$ transforms it into C2. Then, further enhancing the field can drive the system from C2 to C4. Therefore, the FE(C3)-AFE(C1)-FE(C4) phase transition in the polarized state is accompanied by a nontrivial(C3)-trivial(C1)-nontrivial(C4) transition in the topological property. The layer-selective mechanism is not seen for the trilayer and quadlayer, for which the barrier heights for the possible pathways are shown in Appendix Fig.~\ref{fig11}. However, the In$_2$Se$_3$-3L and -4L can still be selectively flipped using the interlocking of the out-of-plane and in-plane polarizations\cite{Cui2018,xiao2018}. Therefore, non-volatile manipulation of topological states can be achieved by switching polarization with appropriate electric fields in In$_2$Se$_3$ film.  

In summary,  we have investigated the ferroelectric and topological properties of In$_2$Se$_3$ thin films. We find that the systems show polarization-dependent electronic structures and band topologies, which can be understood with polarization-dependent band hybridizations. For the In$_2$Se$_3$-2L, the FE states have a nontrivial band topology, whereas others with AFE interlayer couplings are topologically trivial. We also reveal a serials of polarization states with nontrivial band topology for In$_2$Se$_3$-3L and -4L. We find that their topological properties of In$_2$Se$_3$-3L and -4L can be related to that of the In$_2$Se$_3$-2L since they are geometrically connected to the latter. Our results suggest that the In$_2$Se$_3$-2L can be used as a building block for designing architecture with desired band topology.  Thus, for an In$_2$Se$_3$ thin film, the polarization states that having distinct topological properties can be ferroelectrically tunable by making use of the dipole locking effect. Moreover, we have investigated the substrate effect on the topological properties, which remain unchanged upon interfacing with a vdW substrate. While both the rhombohedral (3R, \textit{R3m} ) and hexagonal (2H, \textit{P63/mmc}) In$_2$Se$_3$ thin films were experimentally synthesized for ferroelectricity \cite{Zhou2017,Cui2018, ZengH2018, ZengH2019, LvB2021}, we anticipate that our results will inspire further investigations on the topological properties.

\begin{acknowledgments}
This work was supported by the National Natural Science Foundation of China (Grants No. 12174098, No. 11774084, No. U19A2090,  No. 91833302, and No. 62201208) and project supported by State Key Laboratory of Powder Metallurgy, Central South University, Changsha, China. Calculations were carried out in part using computing resources at the High Performance Computing Platform of Hunan Normal University.
\end{acknowledgments}

%\section{APPENDIX}
%\subsection{Layer projection of the band structure for FE In$_2$Se$_3$-2L and In$_2$Se$_3$-3L}
%\section*{APPENDIX A: \MakeUppercase{Layer projection of the band structure for} In$_2$Se$_3$-2L and In$_2$Se$_3$-3L}

\appendix

\section{Stacking-dependent structural stability for In$_2$Se$_3$-2L}

\begin{figure}[h]
	\includegraphics[width=.5\linewidth]{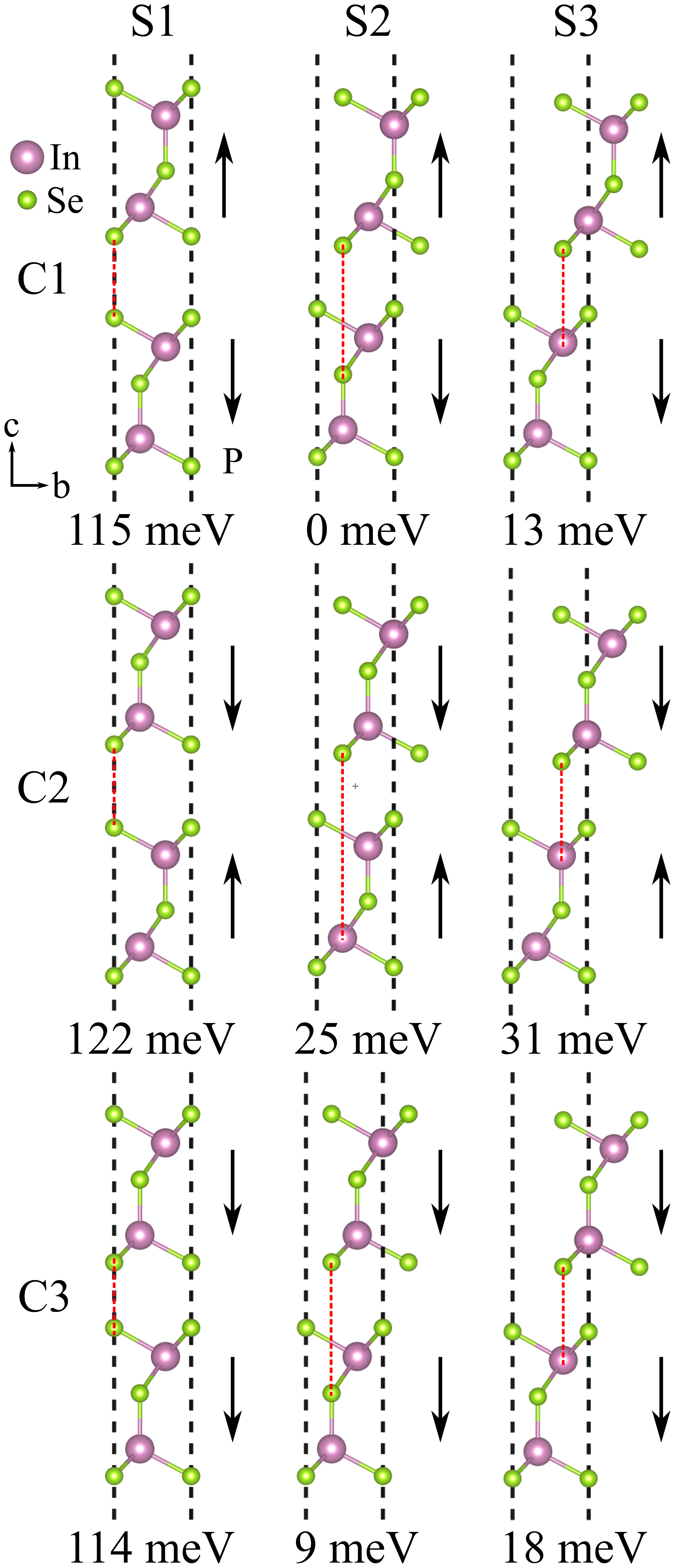}
	\caption{Structures for the three stacking orders of In$_2$Se$_3$-2L. The energy of each structure is given below the structure, for which that of configuration C1-S2 is taken as the reference.}
	\label{fig6}
\end{figure}
Figure~\ref{fig6} shows the considered stacking orders and corresponding energies for all polarization configurations of In$_2$Se$_3$-2L. The upper layer is translated diagonally relative to the bottom layer by 0, (1/3, 1/3), and (2/3, 2/3) for the three stacking orders, respectively. S2 was found to be the most favorable for all polarization configurations. C1-S2 configuration has the lowest energy among all the investigated configurations.

\section{Layer projection of the band structure for FE In$_2$Se$_3$-2L and In$_2$Se$_3$-3L}
%\section{\MakeUppercase{Layer projection of the band structure for} In$_2$Se$_3$-2L \MakeUppercase{and} In$_2$Se$_3$-3L}

\begin{figure}[h]
	\includegraphics[width=.95\linewidth]{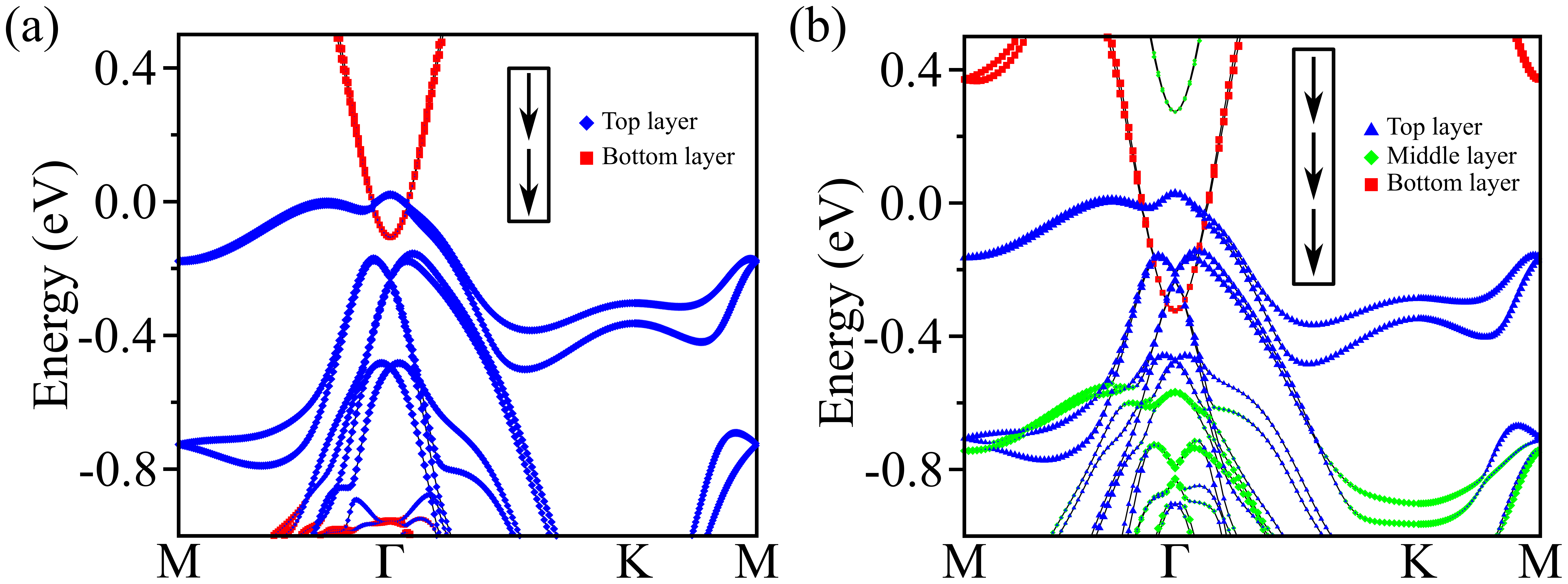}
	\caption{Layer projection of the band structure for FE (a) In$_2$Se$_3$-2L and (b) In$_2$Se$_3$-3L.}
	\label{fig7}
	\end{figure}
Figure~\ref{fig7} shows the layer-projected band structure for FE In$_2$Se$_3$-2L and In$_2$Se$_3$-3L, that is, the band structure is weighted by contributions of different layers.  One can see that the bands near the Fermi level for them are similar. 

\begin{figure*}[t]
	\includegraphics[width=0.9\linewidth]{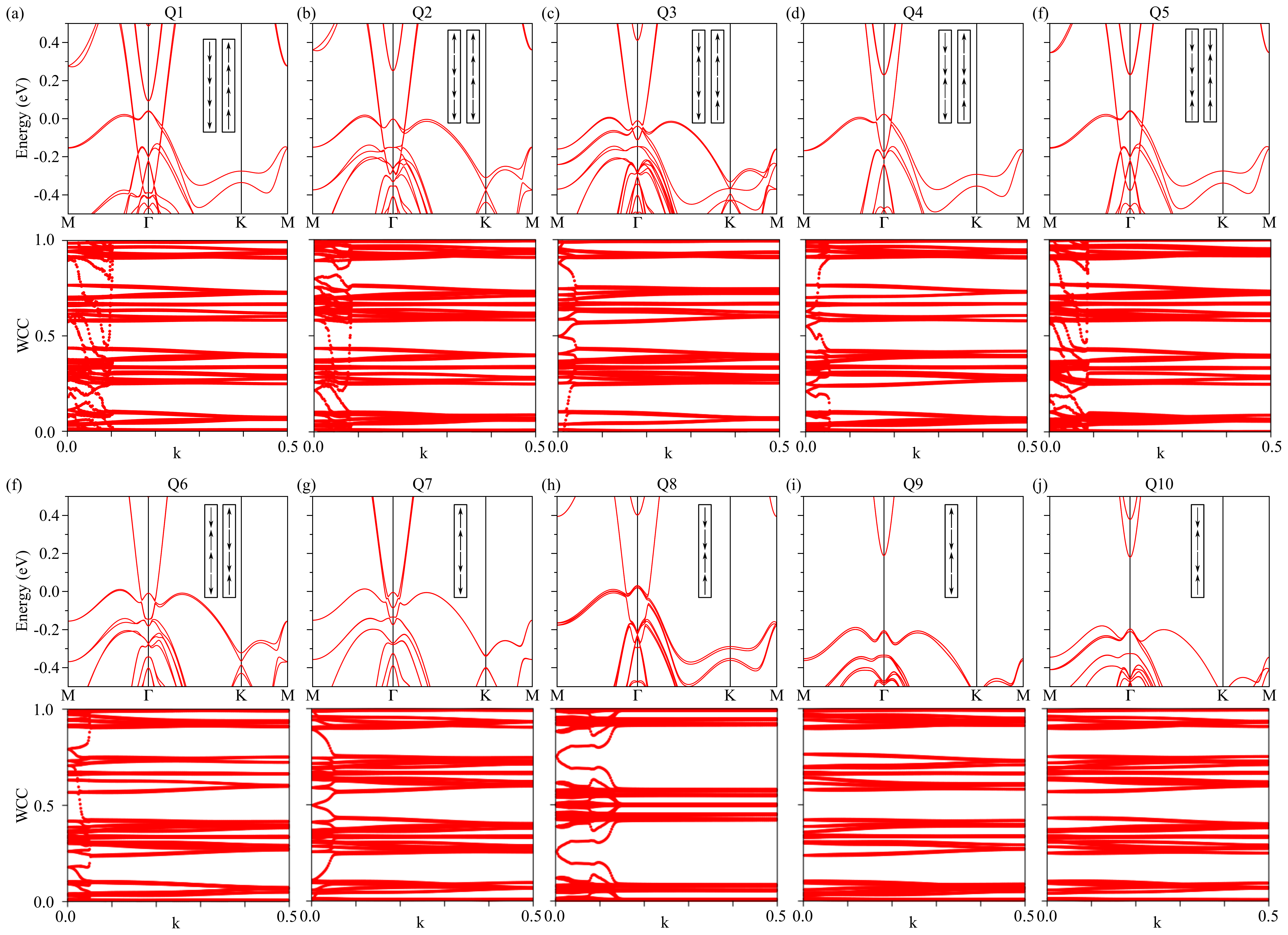}
	\caption{Band structures and evolution of WCC for In$_2$Se$_3$-4L in different polarization states.
	} 
	\label{fig8}
\end{figure*}

\section{Band topology of In$_2$Se$_3$-4L}
Figure~\ref{fig8} shows the band structure and evolution of the WCC for In$_2$Se$_3$-4L in different polarization states, from which one can derive nontrivial topological properties for Q3, Q4, Q6, Q7, and Q8.

\section{Effect of h-BN on the topological properties of FE In$_2$Se$_3$-2L}

\begin{figure*}
	\includegraphics[width=0.8\linewidth]{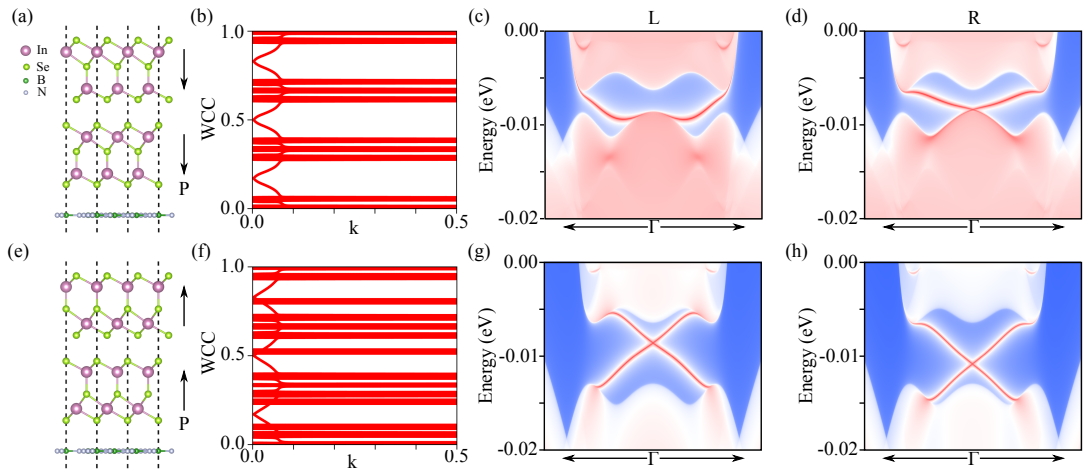}
	\caption{Evolution of WCC and edge states for the FE In$_2$Se$_3$-2L on a h-BN monolayer.
	} 
	\label{fig9}
\end{figure*}

Figure~\ref{fig9} shows the topological properties of In$_2$Se$_3$-2L/h-BN heterojunctions. The results of WCC and edge states demonstrate that the nontrivial band topology for In$_2$Se$_3$-2L is robust against the perturbation of a vdW substrate.

\section{HSE calculations for In$_2$Se$_3$ thin films} 

\begin{figure*}
	\includegraphics[width=.8\linewidth]{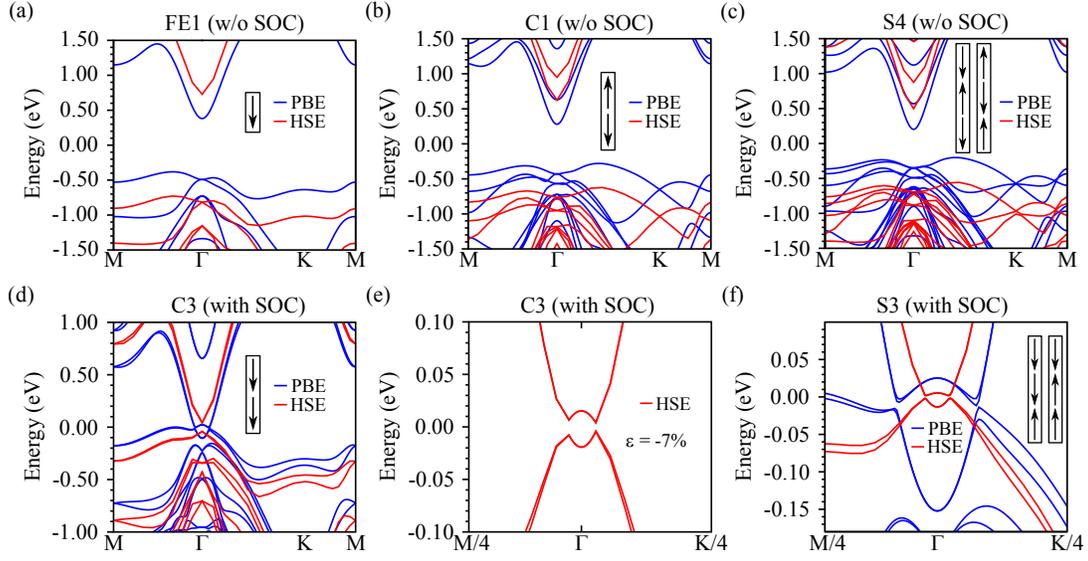}
	\caption{The band structures of the FE1 state for In$_2$Se$_3$-1L (a), the C1 state of In$_2$Se$_3$-2L (b), the C3 state of In$_2$Se$_3$-2L (d), the C3 state with compressive strain $\varepsilon$ of 7$\%$ (e), the S4 states (c), and the S3 states (f) of In$_2$Se$_3$-3L using PBE (blue lines) and HSE (red lines).}
	\label{fig10}
\end{figure*}

We have performed the HSE06 to achieve a more accurate band gap. The band structures calculated using PBE and HSE06 are similar, except for different band gaps (Fig.~\ref{fig10}). Experimentally, the energy band gap for In$_2$Se$_3$-1L is measured to be ~1.55 eV~\cite{Zhou2015}. Figure~\ref{fig10}(a) shows the calculated band gap is 0.60 eV for PBE and 1.46 eV for HSE of In$_2$Se$_3$-1L. In addition, the experimentally measured energy band gap of In$_2$Se$_3$-2L in tail-to-tail
	AFE state (C1) is approximately 1.30 eV\cite{Cui2018}. Figure~\ref{fig10}(b) shows the calculated band gap is 0.56 eV for PBE and 1.25 eV for HSE of In$_2$Se$_3$-2L.These results indicate that HSE06 accurately reproduces the experimental band gap. Furthermore, we have calculated the  band structure for the S3 structure for In$_2$Se$_3$-3L, which is topologically nontrivial. One can see that the SOC-induced band inversion that is the signal of nontrival band topology remains during the HSE06 calculations [Figure~\ref{fig10}(f)].  We have also performed HSE calculations for In$_2$Se$_3$-2L in FE state, which suggest that this system can be topologically nontrivial under appropriate strains.

\section{Kinetic pathway of the ferroelectric phase transform in In$_2$Se$_3$ thin films}

\begin{figure*}
	\includegraphics[width=0.9\linewidth]{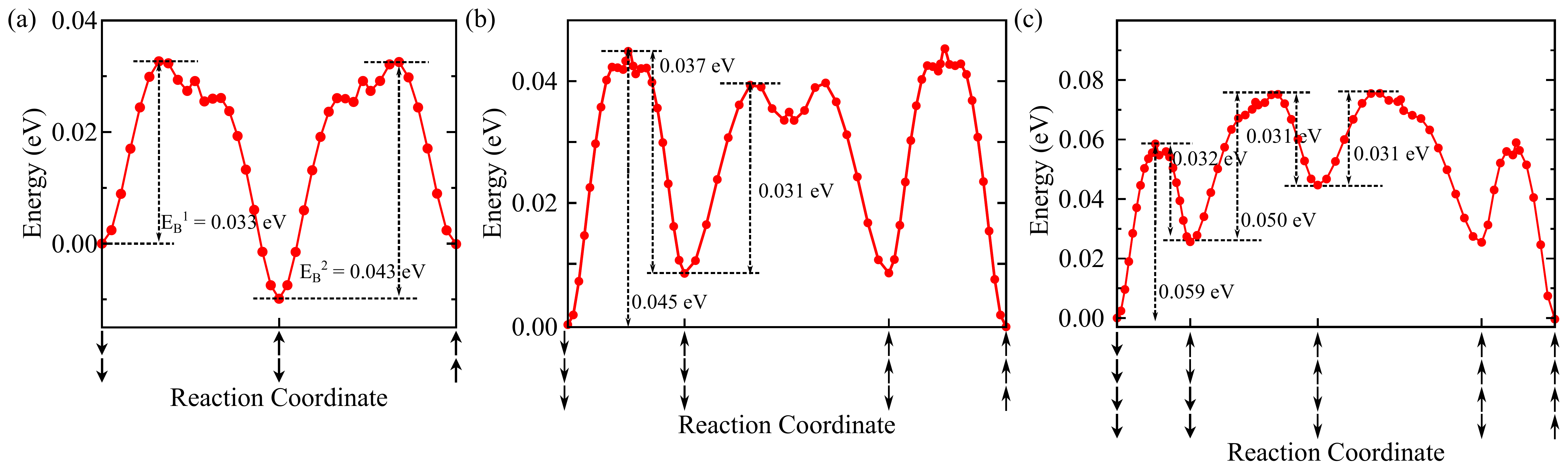}
	\caption{Kinetic pathway of the phase transform in In$_2$Se$_3$ thin films. (a) In$_2$Se$_3$-2L, (b) In$_2$Se$_3$-3L, and (c) In$_2$Se$_3$-4L.
	} 
	\label{fig11}
\end{figure*}
In fact, we calculated the energy barriers for many possible pathways of transforming the polarization states for In$_2$Se$_3$-2L and In$_2$Se$_3$-3L. Figure~\ref{fig11} shows the kinetic pathways of the polarization states transforming in In$_2$Se$_3$ thin films.  For the bilayer, the ground state has an interlayer AFE coupling with tail-to-tail polarizations. As a result, the energy barrier height for the pathway of transforming the system from the FE state to the AFE state is lower than the opposite way.  For the trilayer and quadlayer systems, the ground states are FE. For In$_2$Se$_3$-3L, we calculate the energy barriers by starting off with the FE state, i.e., S1. We obtain the barrier heights by flipping anyone of the In$_2$Se$_3$ layer.  We find that the energy barrier for the surface layer is smaller than those for others. Therefore, the top In$_2$Se$_3$ layer has the priority to be flipped. Then, based on the flipped configuration we proceed to calculate the energy barriers for flipping the middle and the bottom layers. We find that flipping the bottom layer gives rise to a lower barrier height. Finally, starting off with this configuration, we calculate the energy barrier of flipping the middle layer. We use the procedure for the quadlayer.  We also calculate the energy barriers for flipping two In$_2$Se$_3$ layers, which are higher (per In$_2$Se$_3$ layer) than the layer-by-layer flipping.

\section{The calculated PBE band structures using different packages} 

\begin{figure}
	\includegraphics[width=.95\linewidth]{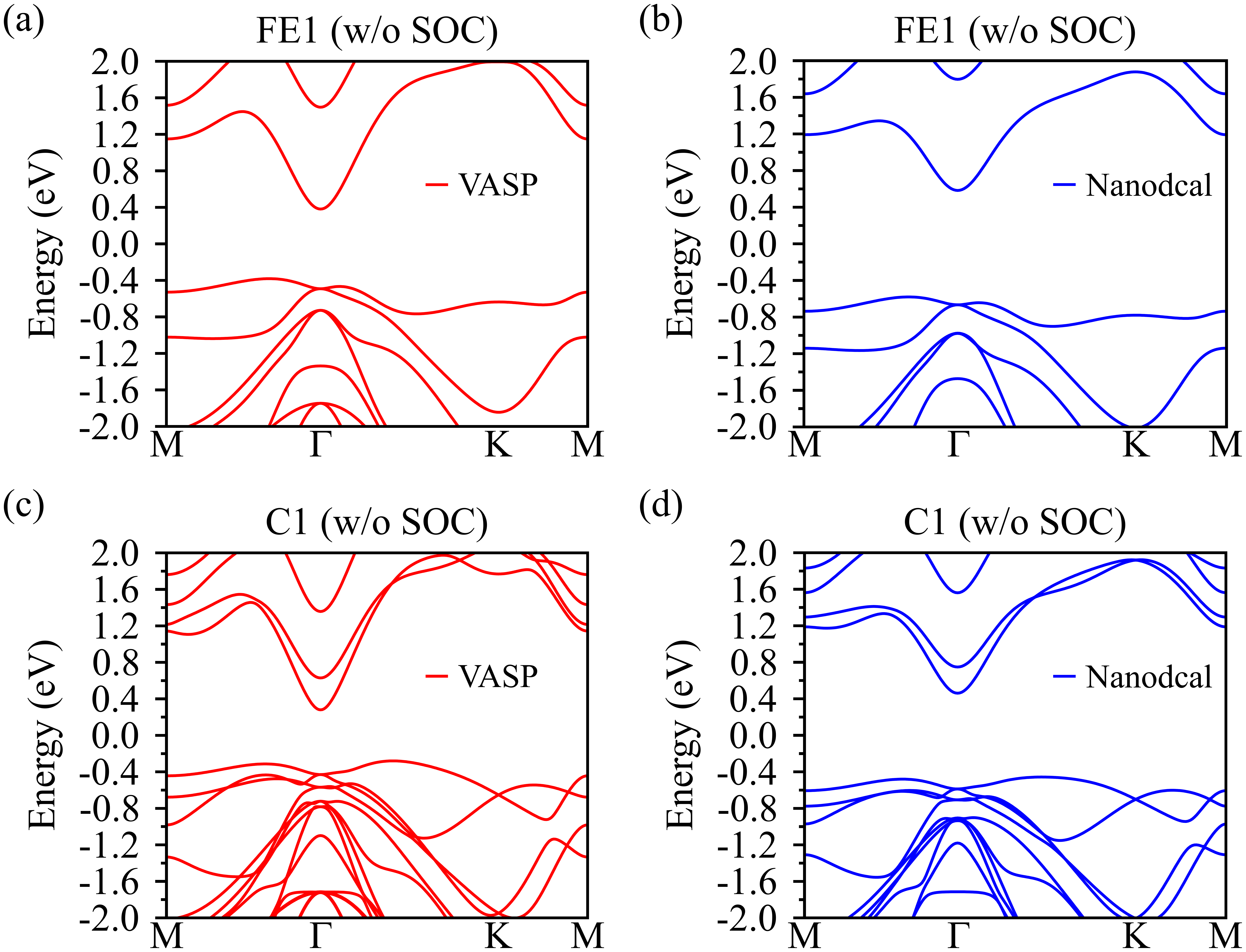}
	\caption{The PBE band structures of the FE1 state for In$_2$Se$_3$-1L (a,b), and the C1 state for In$_2$Se$_3$-2L (c,d). (a) and (c) by using VASP package, (b) and (d) by using Nanodcal package.}
	\label{fig12}
\end{figure}

To ensure the PBE calculation results of VASP, we have performed the Nanodcal packages ~\cite{nanodcal2001} to calculate the band structures of the FE1 state for In$_2$Se$_3$-1L and the C1 state for In$_2$Se$_3$-2L. The corresponding results are shown in Fig.~\ref{fig12}. One can find that the band structures obtained from the two softwares are consistent.

\FloatBarrier

\bibliography{references}
\bibliographystyle{apsrev4-2}

\end{document}